\documentclass[a4paper,11pt]{article}
\usepackage{nfraconf,graphicx,cite}

\usepackage{times}

\begin{document}
\baselineskip=13pt

\title{FREE-FREE ABSORPTION ON PARSEC SCALES IN SEYFERT GALAXIES.}

\author{ A. L. ROY\footnote{Max-Planck-Institut f\"ur Radioastronomie, Auf dem
    H\"ugel 69, D-53121 Bonn, Germany}(aroy@mpifr-bonn.mpg.de)}

\author{ J. S. ULVESTAD\footnote{National Radio Astronomy Observatory,
P.O. Box O, Socorro,  NM  87801, USA}}

\author{ A. S. WILSON\footnote{University of Maryland, and Space Telescope
Science Institute.  Address: Department of Astronomy, University of Maryland,
College Park, MD 20742, USA}}

\author{ E. J. M. COLBERT\footnote{Goddard Space Flight Center,
Mail Code 662, Laboratory for High Energy Astrophysics, NASA Goddard Space
Flight Center, Greenbelt, DM 20771, USA}}

\author{ C. G. MUNDELL\footnote{Department of Astronomy, University of Maryland,
College Park, MD 20742, USA}}

\author{ J. M. WROBEL$^{\mathrm{b}}$}

\author{ R. P. NORRIS\footnote{Australia Telescope National Facility,
P.O. Box 76, Epping NSW 1710, Australia}}

\author{ H. FALCKE$^{\mathrm{a}}$ and T. KRICHBAUM$^{\mathrm{a}}$}
\address{}


\maketitle

\abstract{ Seyfert galaxies come in two main types (types 1 and 2) and the
  difference is probably due to obscuration of the nucleus by a torus of dense
  molecular material.  The inner edge of the torus is expected to be ionized
  by optical and ultraviolet emission from the active nucleus, and will
  radiate direct thermal emission (e.g. NGC 1068) and will cause free-free
  absorption of nuclear radio components viewed through the torus (e.g. Mrk
  231, Mrk 348, NGC 2639).  However, the nuclear radio sources in Seyfert
  galaxies are weak compared to radio galaxies and quasars, demanding high
  sensitivity to study these effects.  We have been making sensitive phase
  referenced VLBI observations at wavelengths between 21 and 2 cm where the
  free-free turnover is expected, looking for parsec-scale absorption and
  emission.  We find that free-free absorption is common (e.g.  in Mrk 348,
  Mrk 231, NGC 2639, NGC 1068) although compact jets are still visible, and
  the inferred density of the absorber agrees with the absorption columns
  inferred from X-ray spectra (Mrk 231, Mrk 348, NGC 2639).  We find one-sided
  parsec-scale jets in Mrk 348 and Mrk 231, and we
  measure low jet speeds (typically $\leq 0.1 c$).  The one-sidedness probably
  is not due to Doppler boosting, but rather is probably free-free absorption.
  Plasma density required to produce the absorption is $N_{\mathrm{e}} \geq 2
  \times 10^5$ cm$^{-3}$ assuming a path length of 0.1 pc, typical of that
  expected at the inner edge of the obscuring torus.  }

\section{Introduction}
Sensitivity of VLBI is constantly improving, and the recent advent of routine
phase referencing with the VLBA has made the imaging of many radio-quiet
active galaxies possible.  These are mostly too weak for self-calibration,
having only a few or tens of mJy core flux density, so we have now been
exploiting phase referencing to make systematic surveys of Seyfert galaxies
with milliarcsecond (mas) resolution, at multiple frequencies to image the
nuclear radio structures and to measure core spectra and jet speeds.

Historically, the torus proposed by Seyfert unification schemes was invoked to
explain why we do not see the broad-line region (BLR) in Sy 2 nuclei that is
thought to be present in most Seyferts.  (The BLR is seen indirectly in some
narrow-line Seyferts through scattering into our line of sight, or by IR
spectroscopy that penetrates the obscuring dust.)  Recently, direct evidence
for the existence of this torus has come from VLBA images of H$_{2}$O maser
discs in NGC 4258 \cite{paper01}, and NGC 1068 \cite{paper02}, and images of
possible thermal emission from the ionized inner edge of the torus in
NGC 1068 \cite{paper03}.  Though masers provide a lot of information (e.g.
$M_{\bullet}$, $L/L_{\mathrm{edd}}$), unfortunately they are rare because
their existence requires special geometry and conditions.  On the other hand,
free-free effects due to the ionized inner edge of an obscuring torus should
be common and should produce characteristic spectral shapes and one-sided jets
and thermal emission that we can look for using multi-frequency VLBI imaging.

\section{Expected Torus Structure}
The torus structure is usually modelled as a slab of gas irradiated
from one side with X-rays from the AGN (e.g.  \cite{paper04,paper05}).  The
predicted temperature and chemical abundances vary with depth into the slab as
the shielding increases, and in a typical model, the
inner edge of the slab lies $\sim 0.3$ pc from the AGN \cite{paper06} and 
the slab is fully ionized for a depth of 0.1 pc at a temperature of
$10{^4}$ K and density of a few$ \times 10^{4}$ cm$^{-3}$.  Such a plasma
would have an optical depth due to free-free absorption of unity at 5 GHz,
or higher if the density is higher or the path length is longer.
Going deeper into the slab the material is warm (6000 K) atomic and then cool
(600 K) molecular, where the maser emission is produced.

Thus, if we view an edge-on torus around an AGN with a core-jet radio
source, the core spectrum should be absorbed and the jet
should be progressively less absorbed along its length like in Cyg A 
\cite{paper07}.

\section{The Surveys}
We selected eight nearby Seyfert galaxies that were known to have strong radio
cores in existing VLA or VLBI images.  We observed three type 1 Seyferts (NGC
4151, NGC 7469, Mrk 231) and five type 2 Seyferts (NGC 1068, NGC 2639, NGC
5506, Mrk 348 and Mrk 463) with the VLBA between 1996 and 1998 at three
frequencies out of 1.6 GHz, 4.8 GHz, 8.4 GHz, and 15 GHz (depending on the
source strength), chosen to span either side of the expected free-free
absorption peak.  The VLBA beamwidth at 1.6 GHz was 5 mas and we tapered the
array at higher frequencies for measuring spectral indices.  We
integrated for 1 h per source per frequency, over a range of
position angles.  We observed a phase referencing calibrator every 2 to 10
min, depending on the frequency, to remove atmospheric and instrumental phase
variations.  Self-calibration then proved able to remove remaining phase
errors and improved the fidelity of the final images.  These were typically
thermal noise limited, at a 5-$\sigma$ level of 2 mJy beam$^{-1}$ in a 2 mas
beam ($35 \times 10{^6}$ K) at full resolution at 5 GHz, or $1.0 \times
10{^6}$ K when tapered to 13 mas resolution for surface brightness
sensitivity.

We have also conducted a search for further examples of thermal emission from
the ionized inner edge of the accretion disc by making deep integrations at
8.4 GHz with the VLBA on five Seyfert galaxies known to contain flat-spectrum
radio cores (T0109-383, NGC~2110, NGC~5252, Mrk~926, and NGC~4388).  The
beamsize was 2 mas, and integration times of 2.5 to 6~h yielded
thermal-noise-limited images with sensitivity of 3-$\sigma = 10^{6}$ K,
sufficient to detect an accretion disc like that in NGC 1068.

\subsection{Results 1: The Core Spectra}

Radio cores in Seyferts typically produce optically-thin synchrotron emission
with $S \sim \nu^{-0.7}$ when imaging with resolution of 100 to 1000 pc
\cite{paper09}.  We found similar results at 0.1 to 1 pc resolution, with most
components showing steep power-law spectra.  However, a number of Seyferts
also showed a flat or absorbed spectrum in at least one nuclear component.  The clear
cases are shown in Fig 1.

\begin{figure}
  \begin{center}
    \vspace{9.8cm}
    \includegraphics{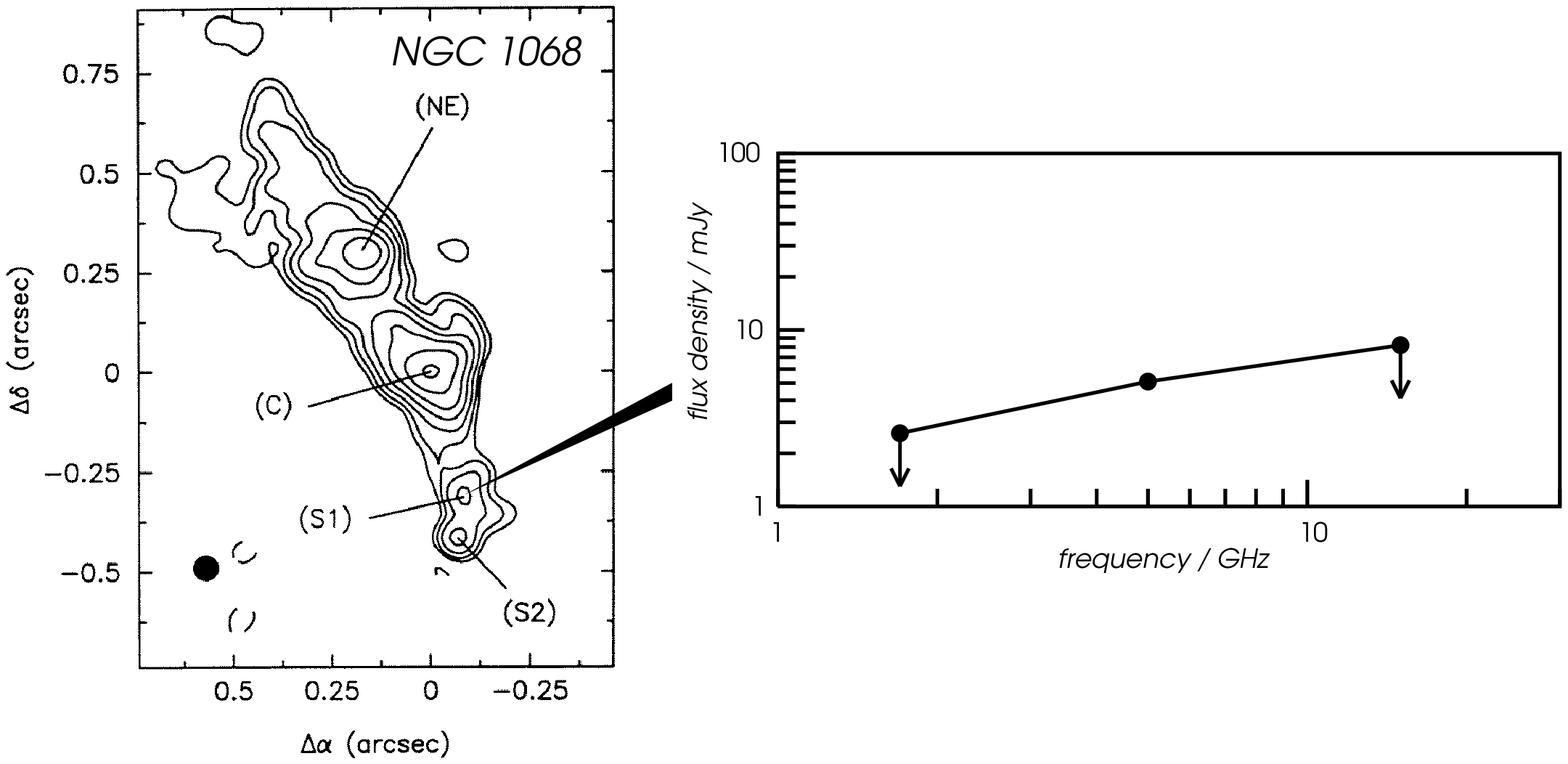}
    \includegraphics{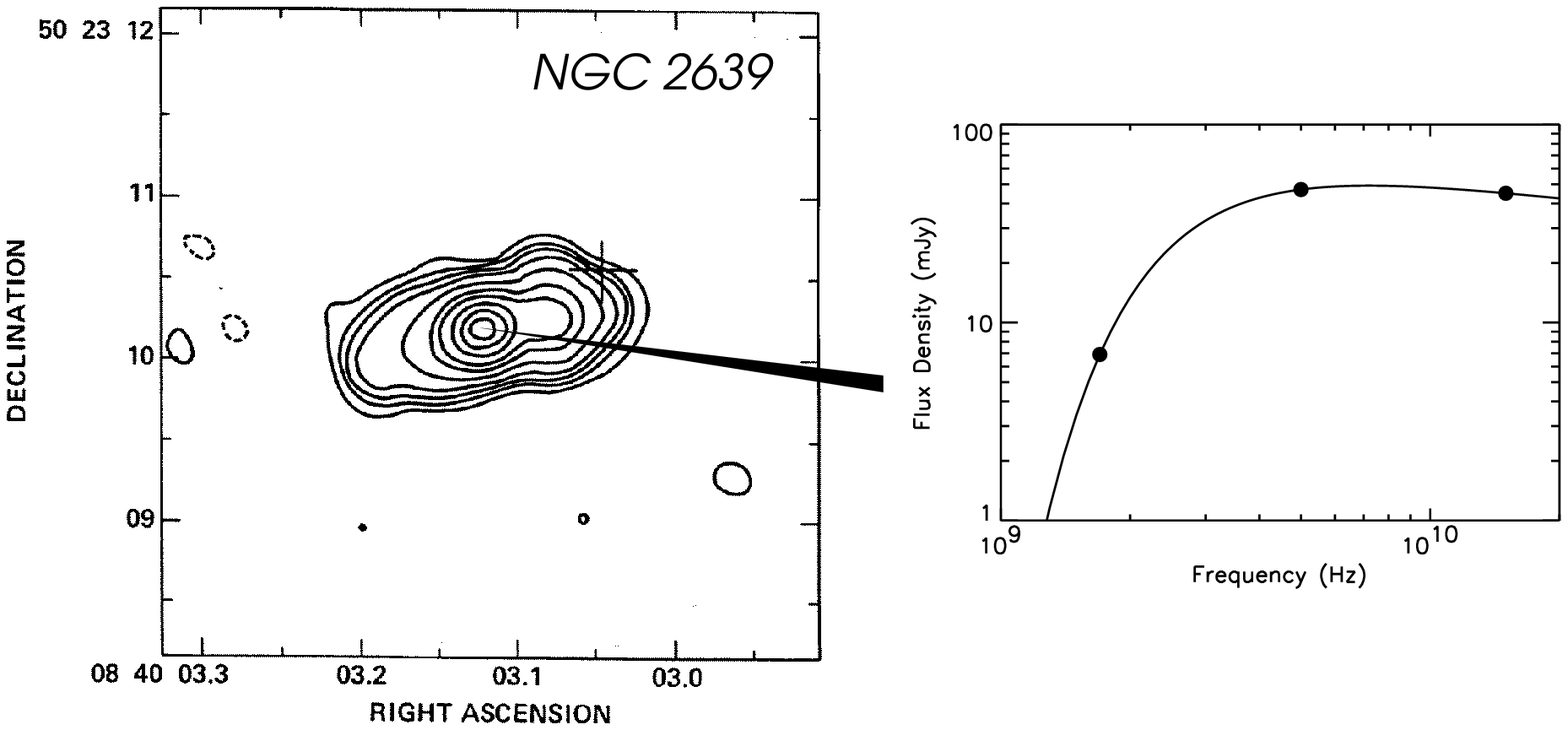}
    \includegraphics{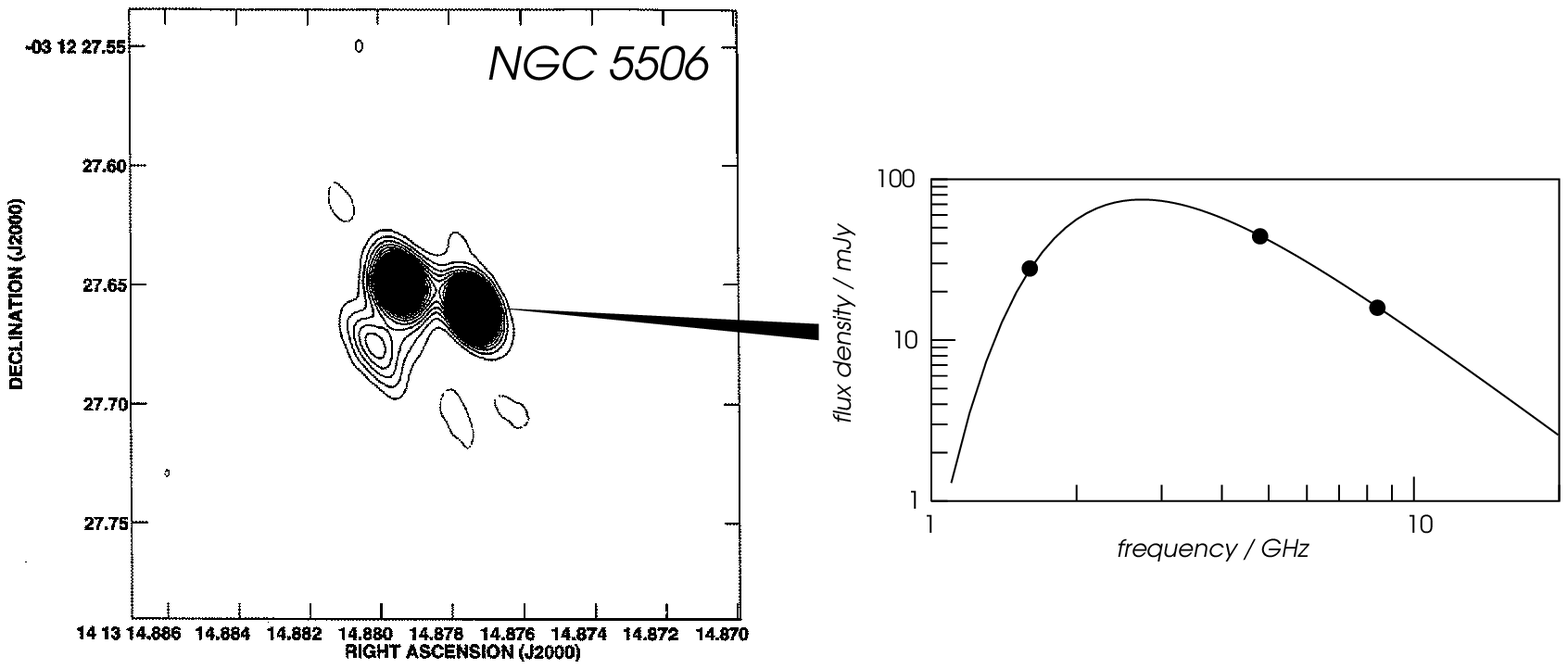}
    \includegraphics{fig1d.ps}
    \includegraphics{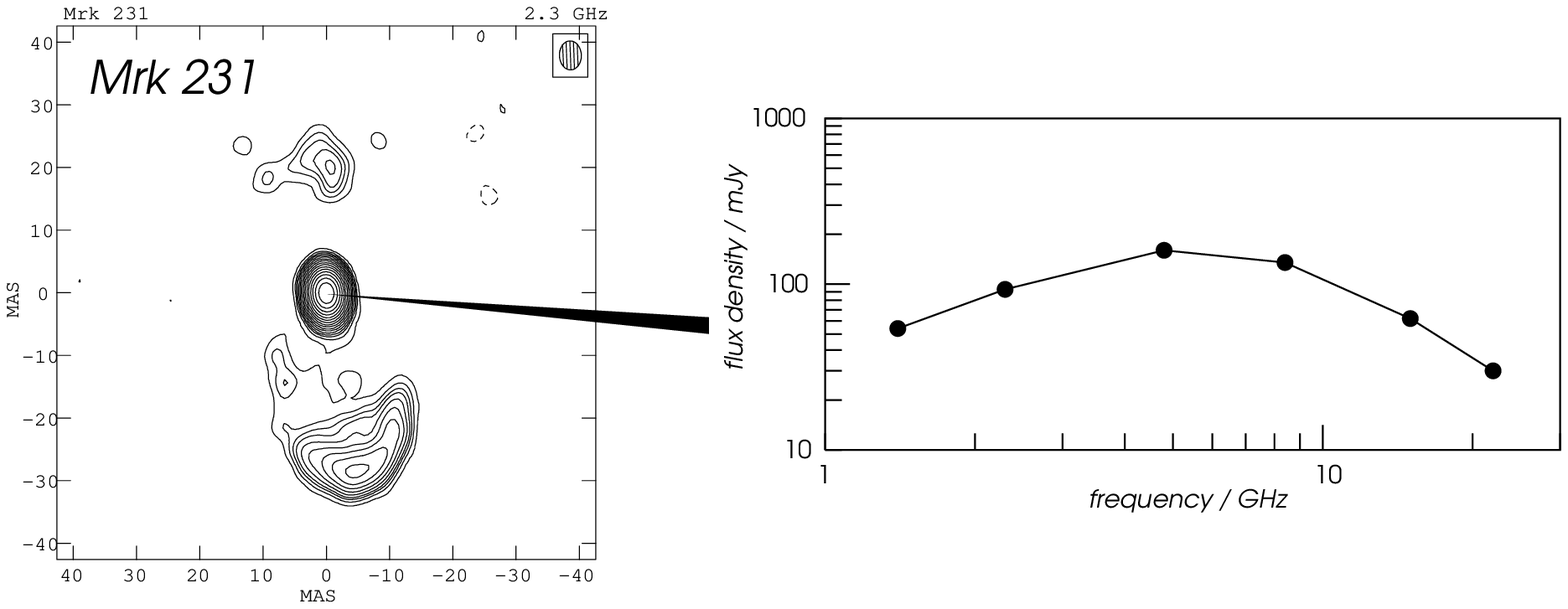}
    \caption{\small The five Seyfert galaxies that show absorbed spectra
      in their nuclear components.  Anticlockwise from top left: NGC 1068,
      MERLIN 6 cm image from \cite{paper10}, and our VLBA spectrum
      \cite{paper11}; NGC 2639, VLA image from \cite{paper09}, and our
      spectrum \cite{paper12}; NGC 5506 image and spectrum from our VLBA
      survey; Mrk 231 image at 13 cm and spectrum from \cite{paper13}; Mrk 348
      imaged by us at 6 cm with a global VLBI array \cite{paper14}, and
      spectrum from \cite{paper15}.}
  \end{center}
\end{figure}

Such absorption could in principle be caused by synchrotron self absorption
(SSA), Razin-Tsytovich suppression, or free-free absorption.  The
Razin-Tsytovich effect is unlikely because it requires field strengths of
$10^{-4}$ gauss for a 0.1-pc absorber, which is 1000 times below
equipartition.

Synchrotron self-absorption requires high brightness temperatures.  For
comparison, in NGC 1068, although the AGN component (S1) has a spectrum of
$\nu^{+0.3}$ around 5 GHz, it is resolved, with $T_{\mathrm{b}}$ of $4 \times
10{^6}$ K \cite{paper03}, which is too low to be caused by SSA (unless
scattering has enlarged the source) and so the rising spectrum is more
naturally accounted for as thermal emission.  In NGC 4151, the nucleus
probably contains a flat-spectrum component, and we find $T_{\mathrm{b}} < 1.1
\times 10^{7}$ K to $> 3.2 \times 10^{7}$ K at 4.8~GHz for the jet and core
components.  The upper range of $T_{\mathrm{b}}$ is a lower limit because the
component E2 is unresolved, so the temperature could in principle be $>
10^{10}$ K, high enough for SSA.  Alternatively, the flat-spectrum
contribution to the nuclear flux could be due to free-free absorption by the
inner part of an HI disc that has been seen in absorption by \cite{paper16}.
NGC 2639 contains an unresolved and variable radio core with an absorbed
spectrum peaking around 5 GHz.  The brightness temperature is $> 1.0 \times
10^{9}$ K at 5~GHz, and so could be SSA.  However, this is also a water maser
galaxy, indicating the presence of an edge-on disc through which we probably
view the radio core.  The inner edge of the disc should be ionized and should
cause free-free absorption of the nuclear continuum as seen.  Mrk 348 and Mrk
231 both have nuclear radio components with absorbed spectra \cite{paper15,
  paper13}.  Brightness temperatures are $> 7 \times 10^{9}$ K (Mrk 348) and
$> 6 \times 10^{9}$ K (Mrk 231), which could be SSA or free-free absorption.
Free-free absorption also could account for the absence of counter-jets (next 
section).

\subsection{Results 2: One-Sided Jets}
Mrk 348 and Mrk 231 both show a core and two-sided ejection on 50-pc scales,
and high-resolution 15-GHz images resolve the cores into 0.5-pc doubles.
Presumably, one component is the core, and the other is a jet on one
side.  A counter-jet is not seen in either object.

The lack of a counter-jet is usually due to Doppler boosting, but here
the jet probably lies across our line of sight \cite{paper17}, and component
speeds are low (below).  Instead, free-free absorption of the counter-jet in
an ionized disc around the AGN is the most natural explanation for the
apparent one-sidedness.

Indeed, the X-ray absorbing column towards Mrk 348
is $10^{23}$ cm$^{-2}$ \cite{paper18}, and towards Mrk 231 is $6 \times
10^{22}$ cm$^{-2}$ \cite{paper19}, which, if distributed over a path of 0.1 pc
like the inner edge of the torus corresponds to densities of $2-3 \times
10^{5}$ cm$^{-3}$, enough to produce free-free absorption of the counter jet.

Interestingly, three Seyferts in our sample are water-maser sources (NGC 1068,
NGC 2639, and NGC~5506) indicating an edge-on disc, and all are Seyfert 2
galaxies and show absorbed core radio spectra, suggesting an absorber on our
line of sight at both optical and radio wavelengths.

\subsection{Results 3: Jet Speeds}
Active galaxies tend to be powerful or weak radio sources and though the
dichotomy has been known for 30 years we still do not know its origin.
Perhaps the engine is the same in both systems and the jet gets disrupted by
dense interstellar medium in the radio-quiet objects (e.g. III Zw 2
\cite{paper20}), or else the difference is intrinsic with jet power scaling
with black hole spin \cite{paper21, paper22}.  To distinguish, one could
measure the jet speed close to the jet base, before
environmental effects have become important.

We have observed Mrk 348 and Mrk 231 at two epochs with a 1.7-yr baseline,
between late-1996/early-1997 and 1998 (Fig 2) \cite{paper17}, and have 18-cm
observations of NGC 4151 at 20-mas resolution with the EVN in 1984 from the
literature \cite{paper08} and with the VLBA at 1996.5.

\begin{figure}
  \begin{center}
    \vspace{6.0cm}
    \includegraphics{fig2a.ps}
    \includegraphics{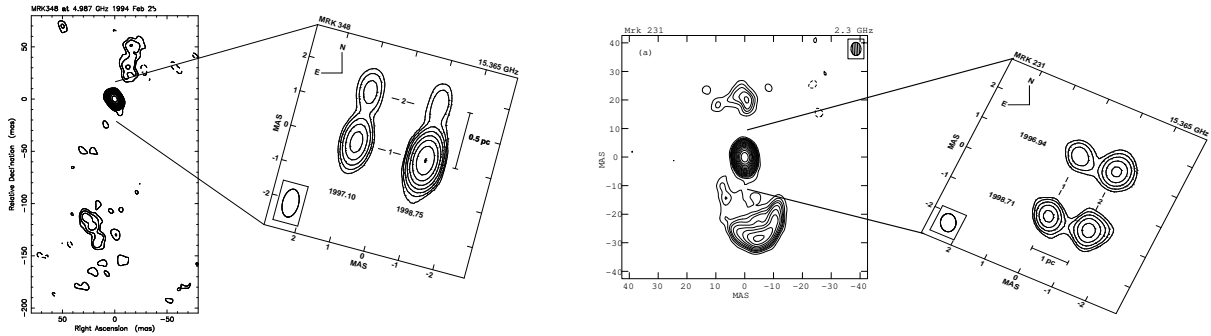}
    \caption{\small Jet motion in two Seyfert galaxies.  Left: Mrk 348 
      imaged at $7.5 \times 5.7$ mas resolution with a Global VLBI array at
      4.8 GHz (\cite{paper14}).  The core is shown at higher resolution
      ($0.80 \times 0.45$ mas) imaged with the VLBA at 15 GHz at two epochs
      (1997.10 and 1998.75 \cite{paper17}).  Right: Mrk 231 imaged at $4.6
      \times $3.5 mas with the VLBA at 2.3 GHz (from \cite{paper13}).  The
      core is shown at higher resolution ($0.7 \times 0.5$ mas) imaged with
      the VLBA at 15~GHz at two epochs (1996.94 and 1998.71 \cite{paper17}).
      Motion between the nuclear components is detectable, yielding
      measurements of the jet speeds of $(0.073 \pm 0.035)$ mas yr$^{-1}$ and
      $(0.046 \pm 0.017)$ mas yr$^{-1}$.  The 15-GHz images are
      self-calibrated and the rms noise is 0.4-0.5 mJy beam$^{-1}$ (except for
      the 2nd-epoch of Mrk 348 which had 1.1~mJy beam$^{-1}$).}
  \end{center}
\end{figure}

The nucleus of Mrk 348 is resolved into a double source whose component
separation increased from 1.46 mas (0.46 pc) to 1.58 mas (0.50 pc) between the
two epochs, yielding a proper motion of $(0.073 \pm 0.035)$ mas yr$^{-1}$.
The nucleus of Mrk 231 is also resolved into a double, whose component
separation increased from 1.08 mas (1.0 pc) to 1.16 mas (1.08 pc), yielding a
proper motion of $(0.046 \pm 0.017)$ mas yr$^{-1}$.  The nucleus of NGC 4151
is resolved into a triple source with component separations of 7 pc and 
36 pc.  No increase in the separations larger than 0.48 pc
and 0.85 pc was found, corresponding to upper limits on the proper motion of 
0.14 and 0.25 $c$.  New VLBA data at higher resolution should soon
measure the jet speed and spectral indices on sub-pc scales in NGC 4151.

In all three objects, the jet speeds are sub-relativistic, different from the
high jet speeds seen in powerful radio sources (Fig 3), and at least two 
are sub-relativistic as they emerge from the broad-line region
(BLR).  This may be due to BLR gas on scales $< 0.5$ pc, or may be 
intrinsic due to black hole spin rate.

\begin{figure}
  \begin{center}
    \vspace{6.0cm}
    \includegraphics{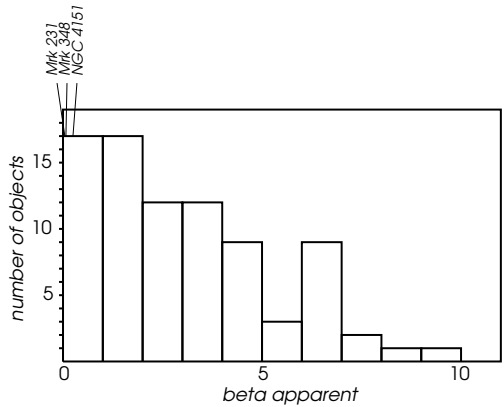}
    \caption{\small The distribution of jet speeds for a uniform sample of
      strong, flat-spectrum objects (comprised of quasars, BL Lac objects,
    galaxies, and empty fields \cite{paper23}), and the speeds measured for
    the Seyferts Mrk 231, Mrk~348, and NGC 4151 shown for comparison.  The
    Seyferts populate the slow end of the distribution.  More Seyferts need 
    to be observed to establish more firmly that their speeds are
    systematically low.}
  \end{center}
\end{figure}

\subsection{Results 4: Misaligned Jets}
In Mrk 231, the parsec-scale jet is misaligned by $65^{\circ}$ with the
40-pc-scale lobes in our 15-GHz VLBA image (Fig 4).  Likewise, in NGC 4151,
the sub-pc-scale jet is misaligned by $65^{\circ}$ with the 300-pc-scale jet
in our 4.8-GHz VLBA image.  The bend in the jet occurs 0.8 pc from the core
(Fig 4).

Bent jets are common in radio-loud objects, but those are probably
projection effects caused by close alignment to our line of sight.  But in
these two Seyfert galaxies, the inclination to the line of sight is probably
large \cite{paper17}, and the bends are probably intrinsic.  A possible cause
might be accretion-disc precession, perhaps by a mechanism as
discussed by \cite{paper25}; timescale for precession may be similar
to that for the jet to propagate the 0.8 pc to the bend at 10000 km s$^{-1}$
\cite{paper24}.  However, the once-only bend requires that such precession
be only transient, and that the nucleus be located at the bend.  The
misalignment complicates standard unification schemes, in which the plane of
the torus defines both the jet axis and the narrow-line
region axis through shadowing, but may be reconciled if the inner disc defines
the jet axis and the outer disc (dust torus) defines the shadowing axis, and
the disc is warped.

\begin{figure}
  \begin{center}
    \vspace{8.0cm}
    \includegraphics{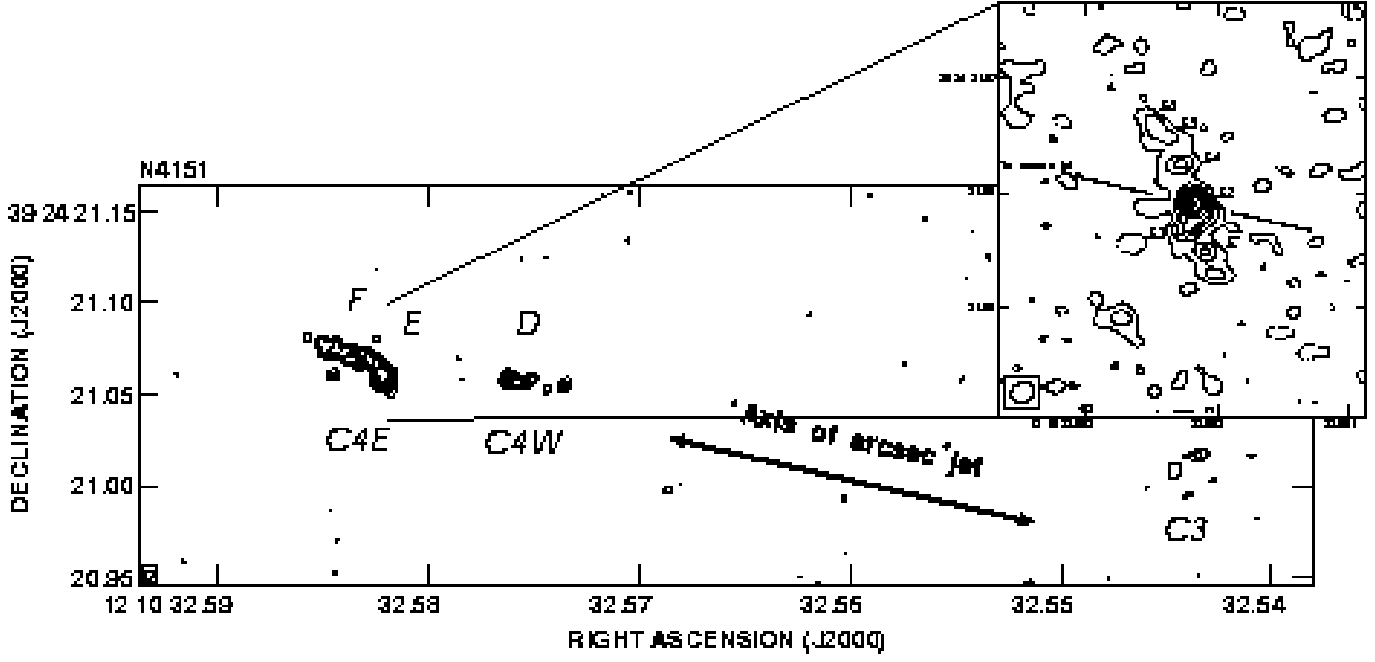}
    \includegraphics{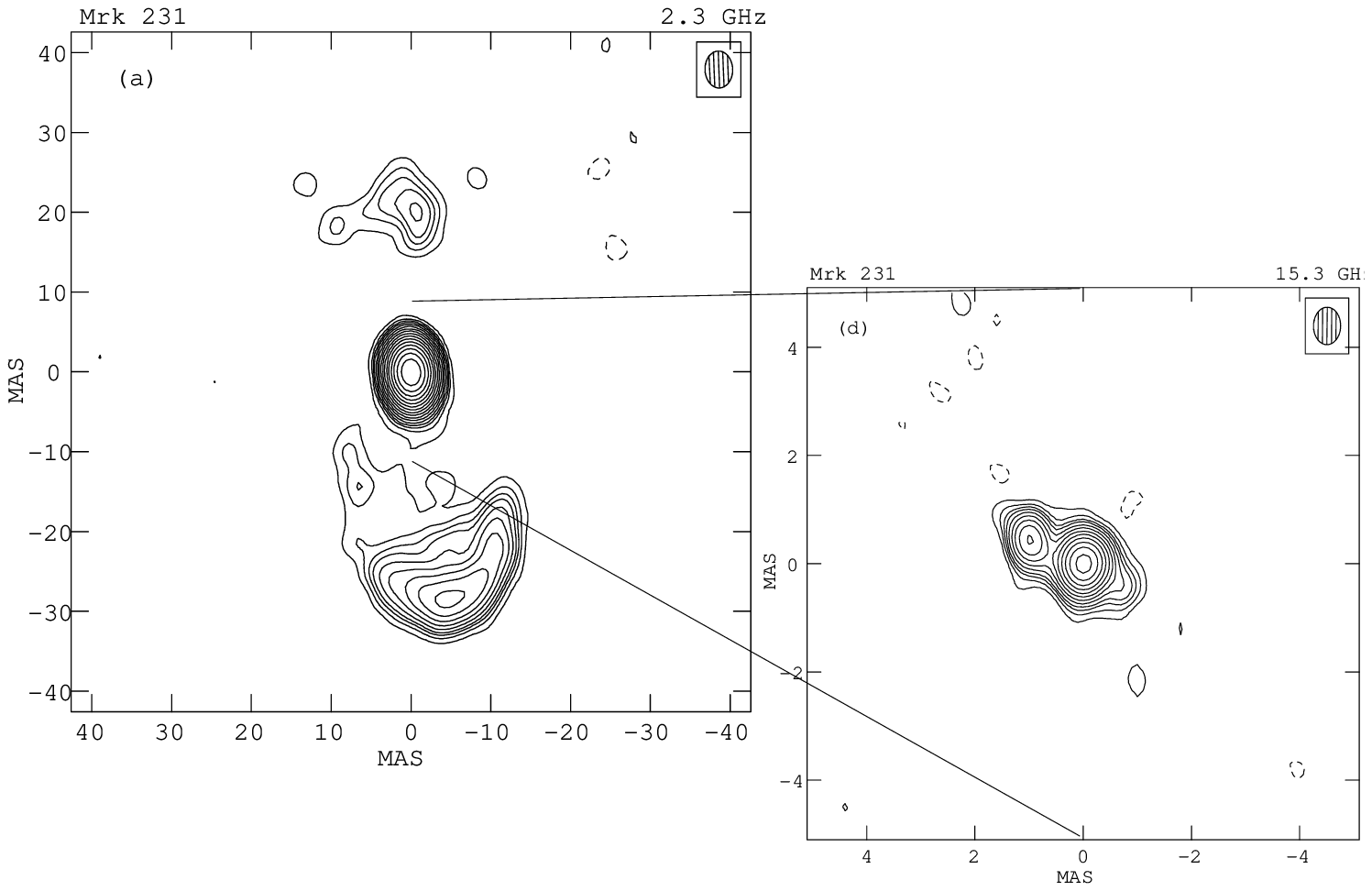}
    \caption{\small Top panel: NGC 4151 VLBA 18 \& 6 cm (inset) images with
      $5.8 \times 4.8$ and $2.1 \times 1.8$ mas resolution \cite{paper24}.
      Bottom panel: Mrk~231 VLBA 13 cm and 2 cm (inset) with $4.6 \times 3.5$
      and $0.7 \times 0.5$ mas resolution \cite{paper13}.  Both objects show
      pc-scale jets that are misaligned with the larger-scale jets.}
  \end{center}
\end{figure}

\subsection{Results 5: Free-Free Emission Search}
We have conducted a search for further examples of thermal emission from the
ionized inner edge of the accretion disc \cite{paper26}.  In the first four
sources (T0109-383, NGC~2110, NGC~5252, Mrk~926) the nuclear source was
detected but was unresolved by the VLBA, indicating $T_{\mathrm{b}}$ in excess
of $10^{8}$ K and sizes less than 1 pc.  The emission is
probably non-thermal and SSA in these galaxies.  In one galaxy (NGC 2110) the
emission was extended by $\sim 0.2$ pc along the direction of the 400-pc scale
VLA jet, suggesting that the emission comes from the base of the jet.  In NGC
4388, the flat-spectrum nuclear source was undetected by the VLBA at 8.4 GHz
but was detected by MERLIN at 5 GHz.  This infers $1.9 \times 10^{4}$ K $<
T_{\mathrm{b}} < 10^{6}$ K, which is too low for SSA and may instead be
optically-thin thermal emission from gas at $> 10^{4.5}$ K and $n_{\mathrm{e}}
> 1.6 \times 10^{4} f^{-0.5}$ cm$^{-3}$, where $f$ is the volume filling
factor.  We are following up with VLBA observations at 1.4 GHz to resolve the
emission region.  Thus, this search found that NGC~4388 may produce thermal
emission from the nucleus, like NGC 1068, and found no evidence for
thermal disc-like emission extended perpendicular to the collimation axis in
the other four galaxies.

\section{Impact of The Square Kilometre Array (SKA)}
The following highlights some Seyfert-related topics on which SKA would make a
big impact.

\subsection{Science}
Thermal emission from the inner edge of the torus may have been seen in
NGC 1068 \cite{paper03}, and a search of five further likely
candidates found another possible example, NGC 4388.  Its
non-detection by the VLBA at 8.4 GHz infers low $T_{\mathrm{b}}$ and we now
need more sensitivity like that of SKA to go further.

Ionized gas in the torus should generate radio recombination lines which, if
detected, would allow one to map the gas dynamics closer to the black hole
than do the water masers, and in many more systems than produce masers.
Surveys for radio recombination lines from Seyfert and starburst galaxies have
already shown that the emission line is weak, being only mJy at cm-wavelengths
\cite{paper27}.  Radio recombination line observations with VLBI will be
challenging and will demand sensitive arrays.

Free-free absorption measurements along the jet and the (presently undetected)
counterjet will allow us to trace the distribution and density of ionized gas
in the inner parsec, revealing whether the absorber is a torus or fills the
nucleus, where the inner edge lies, whether the gas is smooth or clumpy, and
perhaps revealing dense clouds that may be deflecting the jet.  Clearer images
of the jet could show whether the bends are sharp or gradual and show features
due to collisions with dense clouds.

Jet speeds can now be measured for only a few of the brightest Seyferts,
due to sensitivity limits, and speeds seem to be
much less than those in powerful radio galaxies.  To establish this result
more firmly needs larger samples which means monitoring fainter galaxies
than possible at present.

Polarimetry offers measurements of magnetic field strength and structure which
is interesting for studying jet collimation and acceleration in Seyferts.  But
polarized flux is low in the few well-studied Seyferts.  For the strongest,
like NGC 1068, the limits on core polarization are $< 1$\%, but
for most there are no good limits, only non-detections of
polarization.  To go further demands SKA-like sensitivity.  With that, one
might resolve structures transverse to the jets with the highest linear
resolution of any type of AGN, since Seyferts are the the most 
numerous and hence nearest of the AGNs.

\subsection{Imaging Performance}
Weak sources ($< 5$ mJy for VLBA at 6 cm and 128 Mbps) cannot be
self-calibrated because the SNR is $< \sqrt{2}$ on a single baseline within the
atmospheric coherence time.  Phase referencing must be used instead, but,
whilst this permits the detection of weak sources, it incurs a loss of image
fidelity and coherence and causes the appearance of spurious structure due to
residual atmospheric scintillation.  As an example, Fig 5 shows a typical
case.  In this test we observed with the VLBA at 15 GHz two geodetic
calibrators, B1222+037 (410 mJy) and B1226+023 (3C273; 7.6 Jy), both good
point sources, separated by $4.5^{\circ}$, cycling between them with a 5-min
cycle time, for three hours.  After phase referencing from B1226+023 to image
B1222+037, one should get a point source at the phase centre if all worked
well.  However, residual phase errors remain due to small differences in the
atmosphere along the two lines of sight and probably due to errors in the
correlator model.  These cause spurious source structures, loss of coherence,
source scintillation, and time-dependent shifts of the peak position.

\begin{figure}
  \begin{center}
    \vspace{4.5cm}
    \includegraphics{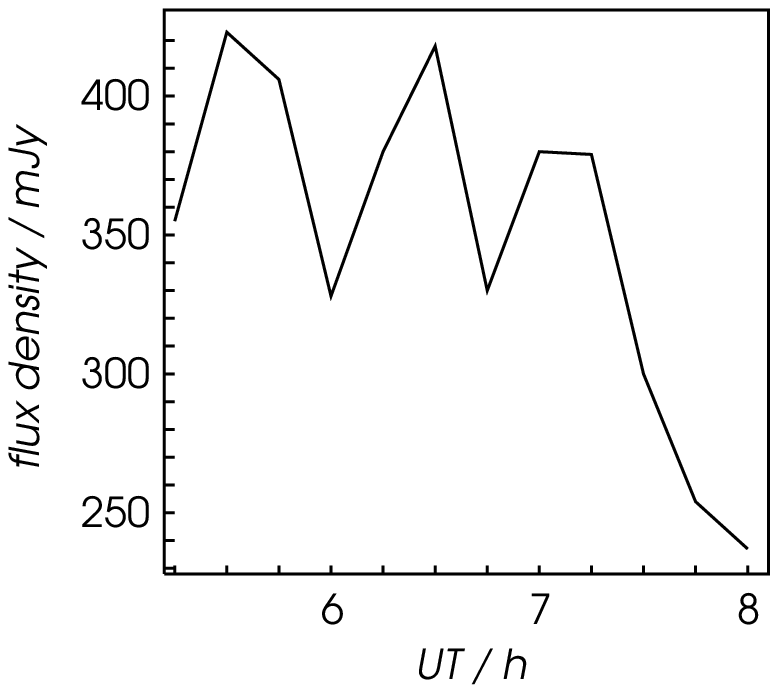}
    \includegraphics{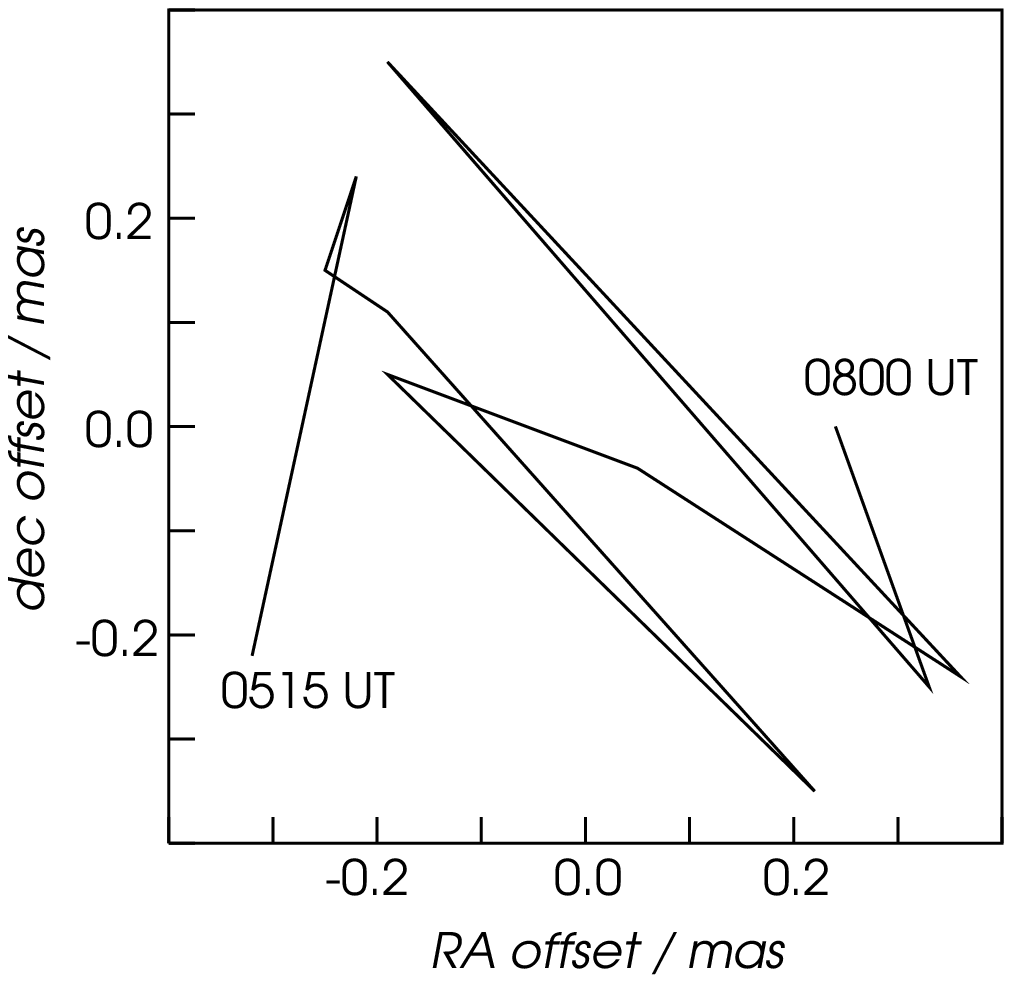}
    \includegraphics{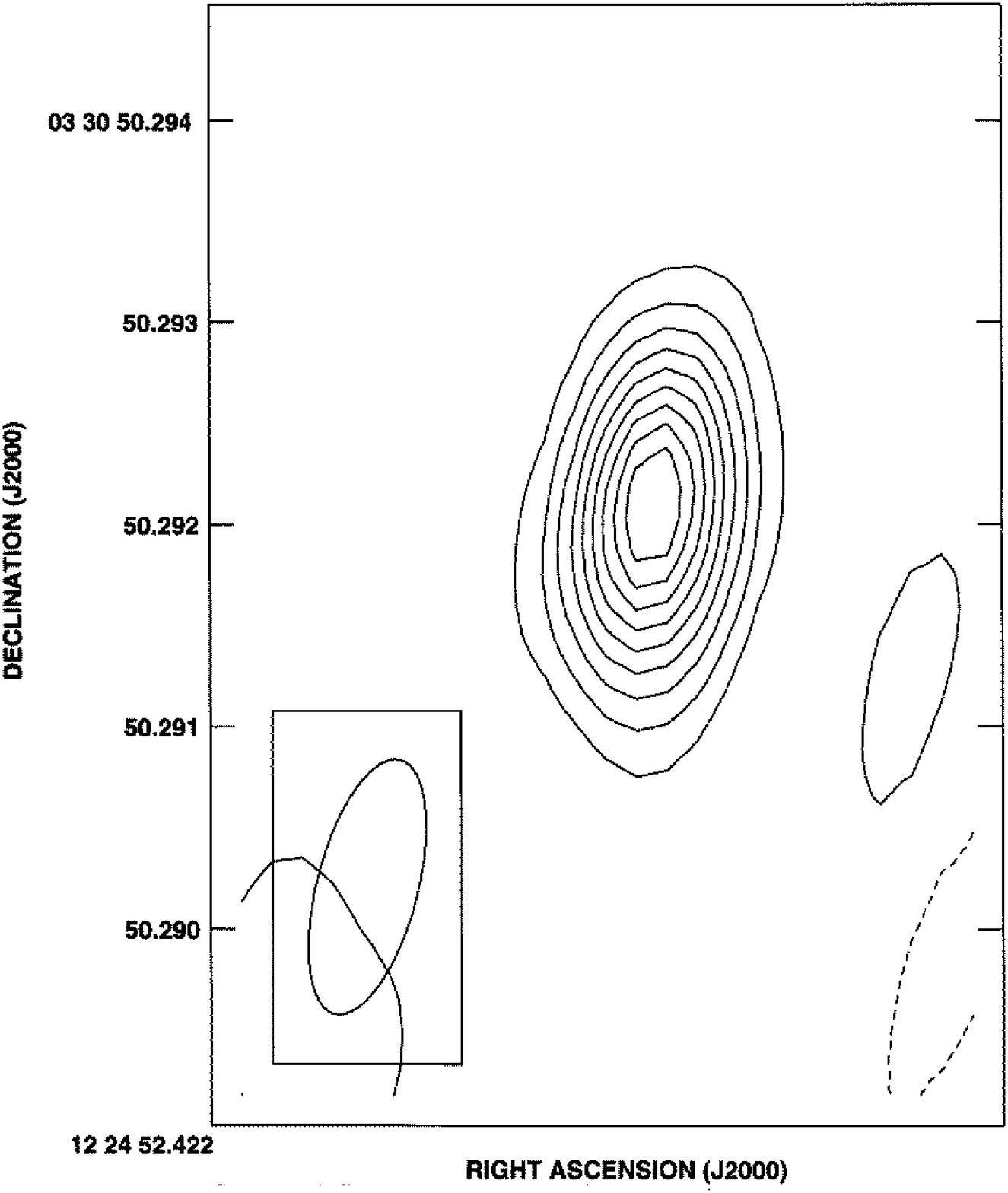}
    \includegraphics{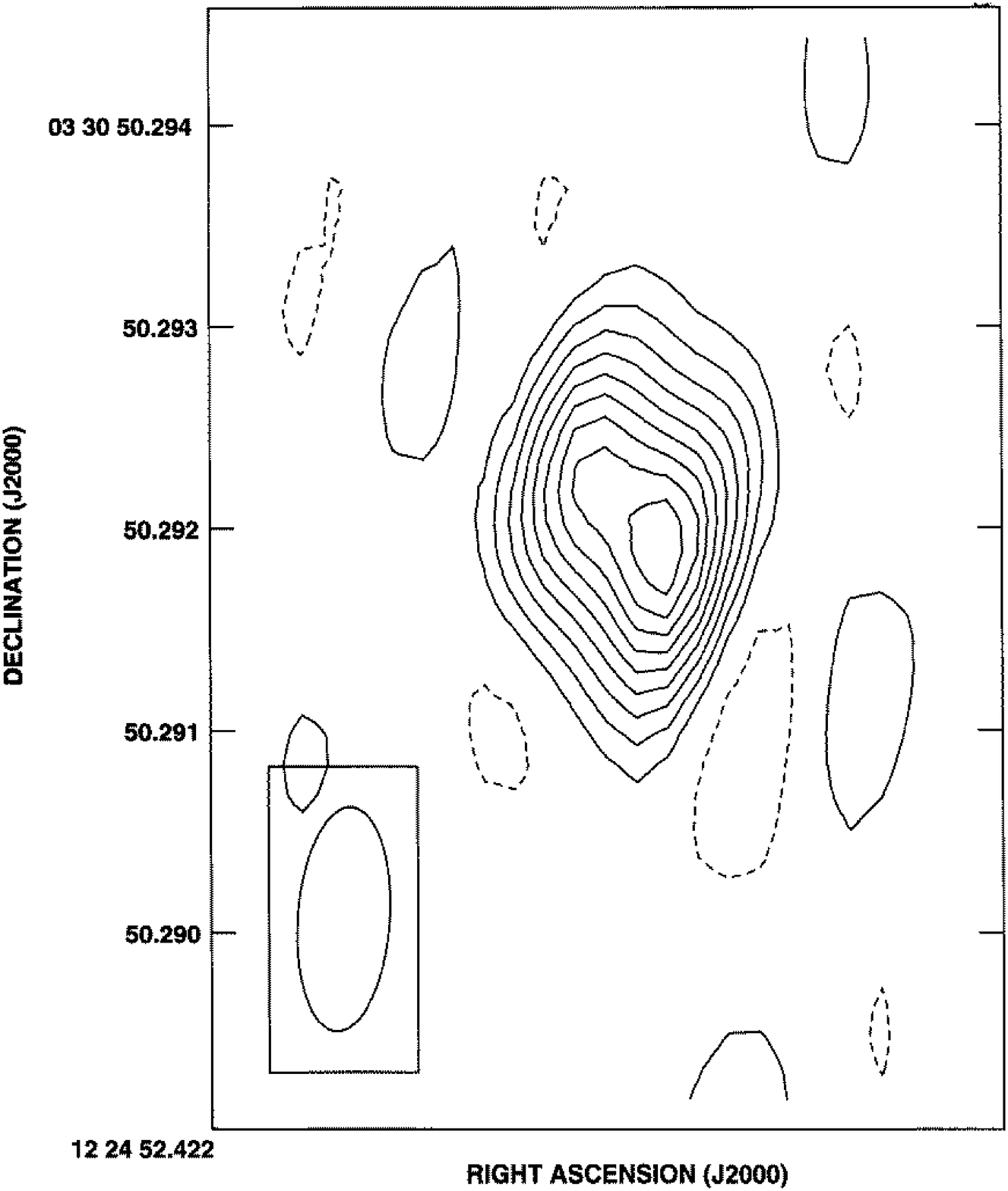}
    \caption{\small
      Test of phase referencing performance.  a) Peak flux density in the
      phase-referenced image of B1222+037 vs time, for images made every 15
      min.  The source is probably constant to $< 1\%$ and the variability
      seen here is due to residual atmospheric scintillation.  b) The position
      of the peak in the phase-referenced image of B1222+037 is shown for
      every 15 min interval throughout the 3-h experiment.  Peak-to-peak
      motion is comparable to the beamwidth ($1.5 \times 0.5$ mas).  c) The
      phase-referenced image of B1222+037 produced from one 15-min snapshot.
      Peak brightness is 424 mJy beam$^{-1}$ and the lowest contour is $10\%$
      of the peak.  d) The phase-referenced image of B1222+037 produced from
      all 3 h of data.  Apparent core-jet structure is caused by source motion
      due to residual atmospheric scintillation.  Peak brightness is 283 mJy
      beam$^{-1}$ and the lowest contour is $10\%$ of the peak.}

  \end{center}
\end{figure}

Much better would be to use the sensitivity of SKA to enable self-cal on
sources down to 0.09~mJy (assumes $\nu = 5$ GHz, $\Delta\nu = 512$ MHz, 1-bit
sampling, $A_{\mathrm{eff}}/T_{\mathrm{sys}} = 9.20$ m$^{2}$ K$^{-1}$ (where
$A_{\mathrm{eff}}$ is the effective area) for each VLBA antenna and
$A_{\mathrm{eff}}/T_{\mathrm{sys}} = 20000$ m$^{2}$ K$^{-1}$ for the SKA,
300-s coherence time, 5-$\sigma$ detection threshold), above which level the
typical source separation is 2", and moderately large VLBI delay-rate beams
can be expected to contain potential calibration sources.

\subsection{Image Sensitivity}
Fig 6 simulates the dramatic improvement in image quality that SKA should
provide.  At the left is a fairly good image of Mrk 348 (15 stations, global
VLBI 1994, 6 cm, 1 h, 0.1 mJy rms), showing a strong core and some weak extended
emission from jets that may be interacting with the narrow-line region gas
\cite{paper14}.  Next is the classic 6-cm VLA image of Cyg A \cite{paper28} to
which has been added noise to degrade it until it `looks' about the same as
the Mrk 348 image.  Last, the noise in the Cyg A image is reduced by a factor
of 50, which is the amount by which the noise would be reduced by adding the
SKA to the VLBA (assumes VLBA-only observes at 6 cm with 128-MHz bandwidth,
1-bit sampling, $A_{\mathrm{eff}}/T_{\mathrm{sys}} = 9.20$ m$^{2}$ K$^{-1}$
per antenna , and VLBA+SKA observes at 6 cm with 512-MHz bandwidth, 1-bit
sampling, $A_{\mathrm{eff}}/T_{\mathrm{sys}} = 9.20$ m$^{2}$ K$^{-1}$ per VLBA
antenna, and $A_{\mathrm{eff}}/T_{\mathrm{sys}} = 20000$ m$^{2}$ K$^{-1}$ for
SKA).  The impact of SKA should be dramatic.

\begin{figure}
  \begin{center}
    \vspace{4.5cm}
    \includegraphics{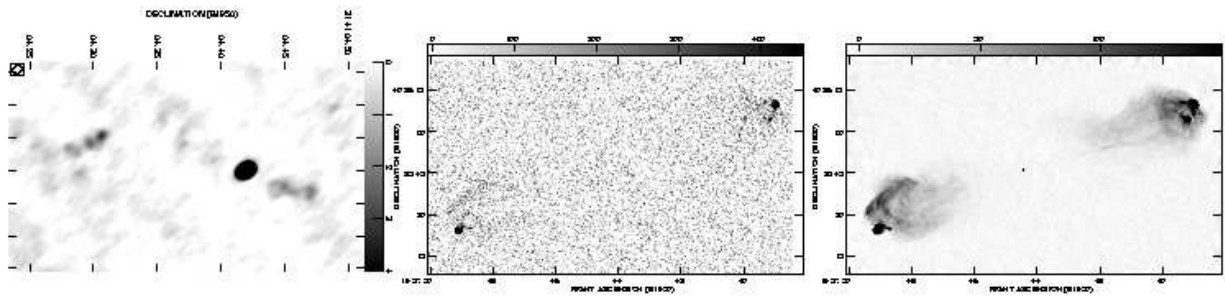}
    \caption{\small
      Left panel: Mrk 348 at 6 cm from a 15-station global VLBI observation
      \cite{paper14}.  Centre panel: Cyg A at 6 cm with the VLA
      \cite{paper28}, with noise added until it `looks' like the Mrk 348
      image.  Right panel: As for the centre panel, but with the noise reduced
      by a factor of 50 as SKA would do if added into an array with the VLBA.}
  \end{center}
\end{figure}

\section*{Acknowledgements}
NRAO is a facility of the National Science
Foundation operated under cooperative agreement by Associated Universities,
Inc.  We thank A.J.Beasley for supplying the phase referencing test data.

\end{document}